\documentclass[letterpaper,10pt]{article}
\usepackage{osameet2}

\usepackage{graphicx}
\usepackage{latexsym}
\usepackage{amsfonts}
\usepackage{amssymb}
\usepackage{psfrag}
\usepackage{cite}
\usepackage{subfigure}
\usepackage{comment,cancel,float}
\usepackage{amsmath,epsfig,epstopdf,color,lineno,makeidx,booktabs,multirow}
\usepackage[pdftex,colorlinks=true,bookmarks=false,citecolor=blue,urlcolor=blue]{hyperref}

\begin{document}

\title{A Robust and Efficient Detection Algorithm for The Photon-Counting Free-Space Optical System}

\author{Tianyu Song and~Pooi-Yuen~Kam}
\address{ Department of Electrical \& Computer Engineering\\ National University of Singapore, Singapore 117583}
\email{\{song.tianyu, elekampy\}@nus.edu.sg}

\begin{abstract}
We propose a Viterbi-type trellis-search algorithm to implement the FSO photon-counting sequence receiver proposed in \cite{Uysal2010GMLSD} more efficiently and a selective-store strategy to overcome the error floor problem observed therein. 
\end{abstract}

\ocis{(060.2605) Free-space optical communication; (040.1880) Detection;}

\section{Introduction}

Since atmospheric turbulence and pointing errors cause fluctuations in the intensity of the received signal,  the photon-counting free space optical (FSO)  system with intensity modulation   requires the channel state information(CSI), i.e., the instantaneous value of the channel gain $h$,  to adjust the detection threshold.
For such a system,  the received signal $r(k)$ in each symbol interval is a discrete Poisson random variable with mean $(n_s m(k) h + n_b)$, where  $n_s$ and $n_b$ are the effective count parameters due to the transmitted signal and the background radiation, respectively  \cite{Uysal2010GMLSD }.
Notation $m(k)$ is used to denote the transmitted data bit, which takes on either the value ``0'' or ``1'' with equal probability.

Since the time scale of the channel fading processes are of the order of $10^{-3}$-$10^{-2}$s, which are far larger than the symbol interval ($\approx 10^{-10}s$ for multi-Gbps systems),
the channel gain $h$ can be regarded as a constant over a large number of symbol intervals and we define this number as the channel coherence length, denoted by $L_c$.   
Thus, at time $k$, the data subsequence with length $L$, denoted by $\mathbf{m}(k,L)=\{m(k-L+1), ... , m(k)\}$, suffers from the same level fading if $L \ll L_c$. 
The authors of \cite{Uysal2010GMLSD} suggested a generalised maximum likelihood sequence detection (GMLSD) receiver, whose structure is
 \begin{align}
 \mathbf{\hat m}(k,L) =  
 \arg \max_{\mathbf{m}(k,L) }  \lambda(\mathbf{m}(k,L)).
\label{eq:receiver_gmlsd_seq} 
 \end{align}
In \eqref{eq:receiver_gmlsd_seq} , the decision metric  is
\begin{align}
 \label{eq:metric_gmlsd_seq}
 \lambda(\mathbf{m}(k,L)) = \left[\frac{R_\mathrm{on} }{N_\mathrm{on} n_b} \right]^{R_\mathrm{on} }   \exp[-R_\mathrm{on} +n_bN_\mathrm{on} ],
\end{align} 
 where $R_\text{on}$ and $N_\text{on}$ are functions of $\mathbf{m}(k,L)$ and are defined as
 \begin{align}
N_\mathrm{on}(\mathbf{  m}(k,L))=\sum_{i=k-L+1}^k m(i), \qquad\qquad 
R_\mathrm{on}(\mathbf{  m}(k,L))=\sum_{i=k-L+1}^k m(i)r(i).
\label{eq:Ron_def}
\end{align}
Evidently, the decision metric of this GMLSD receiver given in \eqref{eq:metric_gmlsd_seq} avoids the calculation of complex integrals and thus can be evaluated efficiently.
Moreover, this decision metric is based solely on the observed values of received subsequences.
No more channel model information is required and this makes it practically implementable. 
As discussed in \cite{Uysal2010GMLSD}, since it can simultaneously estimate the channel and detects the data sequence,  it can approach the Genie Bound, which is defined as the bit error probability (BEP) of the ideal receiver given in \cite[Eq. (9)]{Uysal2010GMLSD}, with a large observation window size.

As introduced in \cite{Uysal2010GMLSD}, the multiple-symbol detection (MSD) algorithm proposed in \cite{Schober2008PCTWC} is adopted to implement the sequence receiver \eqref{eq:receiver_gmlsd_seq}. 
The MSD algorithm employs a block-by-block detection scheme, and the search complexity is $\text{O}(L\log L)$.
In general, a higher $L$ is preferred since it can help the receiver to achieve a better performance. 
However, if $L$ increases, the search complexity increases and also a large delay is introduced.  
In addition, an error floor problem is observed affecting the receiver performance. 
In the next section, we propose our robust and efficient implementation method of this GMLSD receiver.

\section{The Viterbi-Type Trellis-Search Algorithm, the Selective-Store Strategy and the Performance}
 
\begin{figure}[htbp]
\centering 
\subfigure[Trellis]{
\includegraphics[scale=0.9]{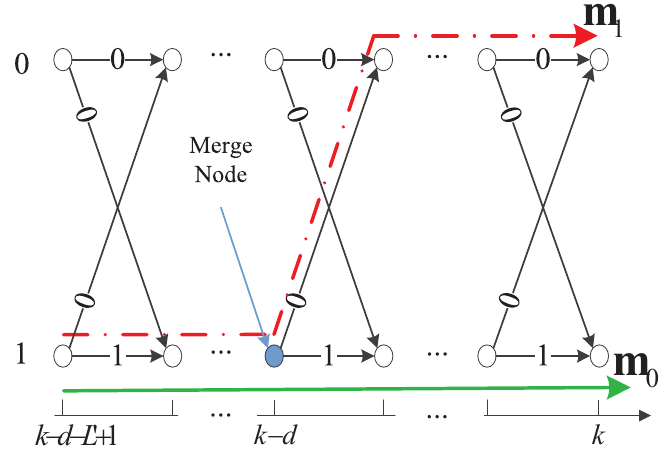}%
\label{fg:GLRT_trellis}
}
\subfigure[Memory Usage]{ 
\includegraphics[scale=0.9]{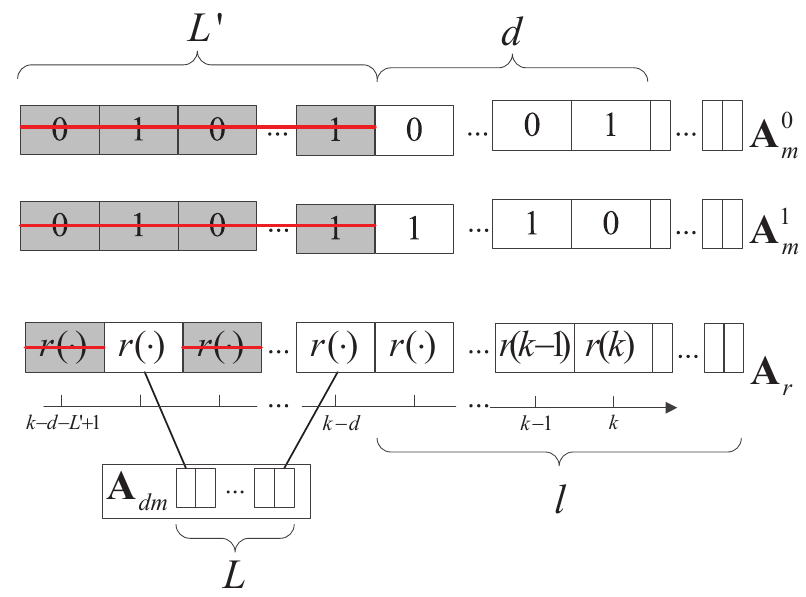}%
\label{fg:memory_gmlsd_seq}
}
\caption{The trellis diagram and the memory usage diagram}
\end{figure}

Similar to \cite{Song2014Arobust}, as shown in Fig. \ref{fg:GLRT_trellis}, there are 2 nodes at each time point in the trellis diagram. 
We label each node using values assumed by the data ``0'' or ``1''. 
For any paths, entering the same node represents the transmission of the symbol corresponding to that node. 
At this point, we evaluate the decision metric of a branch with only its corresponding $L$ most recent symbol vector $\mathbf m(k,L)$ and the $L$ most recent received signals $\mathbf r(k,L)$.  
Any two paths entering the same node are compared in terms of their metric value and   the one with a lower metric value is discarded.
In this way, the path with the largest metric is saved as the survivor. 
The decision on a bit is made only when the tails of all survivors have merged at the corresponding node. 
For example, as shown in Fig. \ref{fg:GLRT_trellis}, two paths, $\mathbf {m}_0(k,L)$ and $\mathbf {m}_1(k,L)$ merge at time $k-d$. 
Accordingly, all the decisions before time $k-d$ are made.
The entire trellis-search algorithm works in the same way as the Viterbi algorithm.
Clearly, since for each time point $k$, only four possible decision metrics need to be evaluated and two of them with smaller metric values are discarded, the overall search complexity of our Viterbi-type trellis-search algorithm is small and is independent of the observation window length.

An error floor problem is observed in \cite{Uysal2010GMLSD}. 
This is because when a  block incurs all zero symbols, the denominator of the decision metric becomes zero since $N_\text{on}$ is zero. 
In our work, we adopt a selective-store strategy to overcome this problem.
In advance, we define the received signal $r(k)$, which is detected to carry symbol 0, as the 0-detected signal; and similarly, the one that is detected to carry symbol 1 is defined as the 1-detected signal. 
We can see that each of the two survivors showed in Fig. \ref{fg:GLRT_trellis} can be divided into two parts: the detected part and the ongoing part.   
For the case shown in Fig. \ref{fg:GLRT_trellis}, the detected part is the one before (including) time $k-d$; and the ongoing  part is the one after time $k-d$.  
As we can see from \eqref{eq:metric_gmlsd_seq} that the 0-detected signals are not contributing to the decision metric \eqref{eq:metric_gmlsd_seq}, for the detected part, we need only store the most recent $L_m$ 1-detected signals, where $L_m$ is used to denote the memory length. 
We define a memory array $\mathbf{A}_{dm}$ with length $L_m$, as shown in Fig. \ref{fg:memory_gmlsd_seq}, to store the most recent $L_m$ 1-detected signals. 
In this way, with this selective-store strategy, the receiver has sufficient information for estimating the unknown channel. 
In addition, we need two more memory arrays to store the ongoing parts of the two survivors and one more array to store the ongoing received subsequence.
As shown in Fig. \ref{fg:memory_gmlsd_seq}, we define them as $\mathbf{A}^0_m$, $\mathbf{A}^1_m$ and $\mathbf{A}_r$ with length $l$. 
We use $L'$ to denote the real length of the detected part that contains exactly $L_m$ 1-detected signals and $d$ to denote the length of the ongoing part. 
Apparently, the total subsequence length is $L=L'+d$ and is a random variable since $L'$ and $d$ are both random due to the transmit data uncertainty. 
From simulation observations, the average value of $d$ is no larger than 2, thus $L$ is of the order of $2L_m$.
We have to choose an appropriate $L_m$ value to ensure $L \ll L_c$. 
Also, to ensure $d \leq l$, we set $l=20$.
Using this selective-store strategy, we selectively store the the most recent 1-detected signals, we thus prevent the denominator of the decision metric \eqref{eq:metric_gmlsd_seq} being zero, and consequently overcome the error floor completely.  
Besides, the delay of the system using our algorithm depends on the value of $d$.
Since $d$ is always very small,  the system delay is very small (generally less than 2 bits) and is independent of $L_m$, or $L$.

\section{Numerical Results and Discussion}

\begin{figure}[htbp] 
\subfigure[Log-normal model, S.I. = 0.5.]{
\includegraphics[scale=0.93]{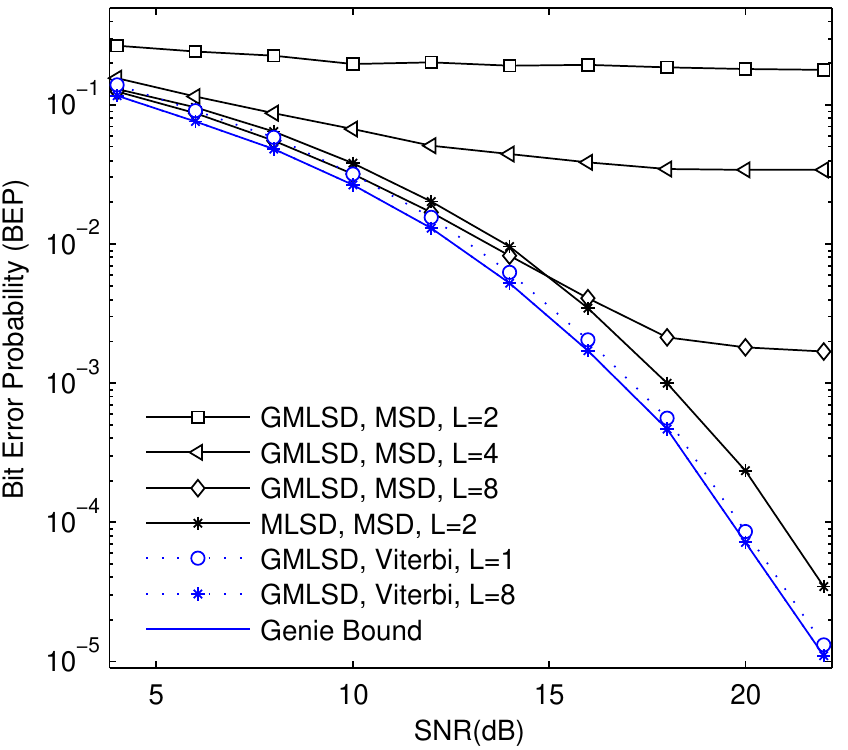}
\label{fg:compare_fig1}
}
\subfigure[  Gamma-Gamma  model, S.I. = 1.38.]{
\includegraphics[scale=0.93]{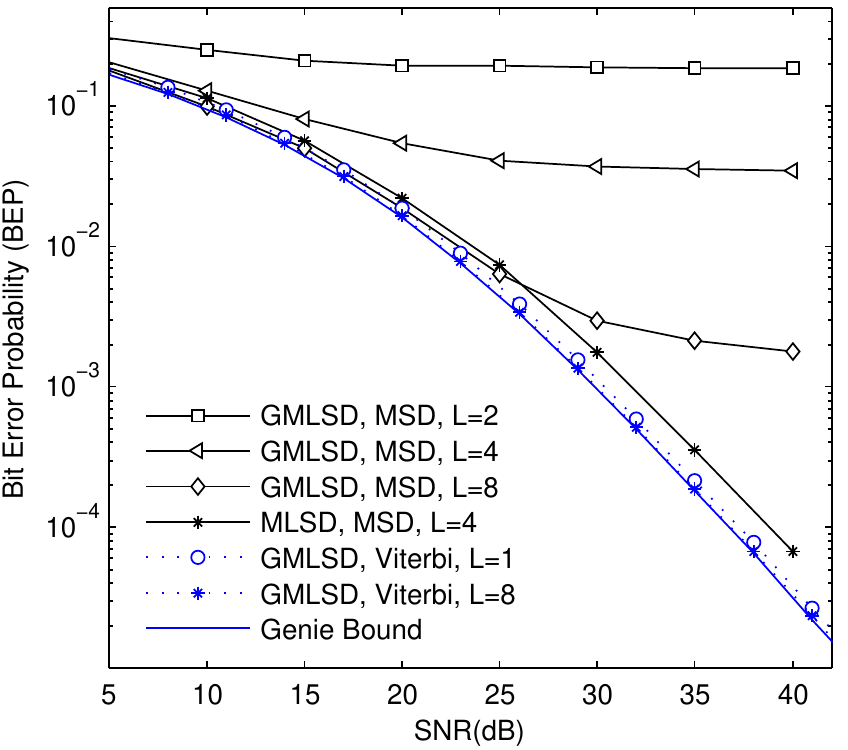}
\label{fg:compare_fig2}
}
\vspace{-6pt}
\caption{  Comparison of the MSD Algorithm and the Viterbi-type Trellis Search Algorithm}
\label{fg:compare}
\vspace{-20pt}
\end{figure}

We plot Fig. \ref{fg:compare_fig1} corresponding to  \cite[Fig. 1]{Uysal2010GMLSD},  Fig. \ref{fg:compare_fig2} corresponding to \cite[Fig. 2]{Uysal2010GMLSD}. 
Both the channel models and model parameters adopted for Fig. \ref{fg:compare_fig1} are the same as that for \cite[Fig. 1]{Uysal2010GMLSD} and that for \ref{fg:compare_fig2} are the same as that for \cite[Fig. 2]{Uysal2010GMLSD}.
The definition of signal-to-noise ratio (SNR) used in this paper is the same as that in   \cite{Uysal2010GMLSD} and is given in  \cite[eq. (7)]{Uysal2010GMLSD}; and the definition of scintillation index (S.I.) used here is given in \cite[eq. (1)]{Uysal2010GMLSD}.
We do not re-implement the GMLSD receiver using the MSD algorithm, and when re-plotting curves in \cite{Uysal2010GMLSD} for comparison, we precisely read   \cite[Fig.'s 1 and 2]{Uysal2010GMLSD} by Adobe Illustrator to obtain the data values (BEP versus SNR).   
For systems using the MSD algorithm, $L$ is used to denote the block length and for systems using our algorithm, $L_m$ is used to denote the memory length.

Clearly, we can see that with $L_m=1$, the performance of our implementation is very close to the Genie Bound. 
It outperforms the GMLSD receiver using MSD algorithm with $L=2$, $L=4$ and $L=8$ and it is even better than the MLSD receiver (with $L=2$) proposed in \cite{Schober2008PCTWC} which requires the channel model information.
Though the difference between the performance of our implementation method and the Genie Bound can be observed.
We are achieving the Genie Bound by increasing the value of $L$.
When $L_m$ increases to 8, the performance of our implementation has perfectly achieved the Genie Bound. 
For the GMLSD receiver implemented by the MSD algorithm, to achieve the Genie Bound at larger-than-$10^{-5}$ region, $L$ is required to be no smaller than 50.  
More importantly, from Fig. \ref{fg:compare}, no error floor has been observed even with $L=1$.

\section{Conclusion}
In this paper, we investigate implementation methods of the GMLSD sequence receiver for the FSO photon-counting communication system. 
We propose a Viterbi-type trellis-search algorithm  to improve the search efficiency, and a selective-store strategy to overcome the error floor problem as well as to increase the memory efficiency. 
Using simulations, we have shown the performance of our implementation method outperforms that with the MSD algorithm and can approach the Genie Bound with a relatively small observation window size. 
Additionally, adopting our method can help completely avoid the error floor problem.

\end{document}